\documentclass[floatfix,aps,amsmath,nofootinbib,twocolumn,10pt]{revtex4}

\usepackage{listings}
\usepackage{graphicx}
\usepackage{bm}
\usepackage{rotating}
\usepackage{array}
\usepackage{amsmath}
\usepackage{amssymb} 
\usepackage{mathrsfs} 
\usepackage{cancel}
\usepackage{diagbox}
\usepackage{color}

\def\({\left(}
\def\){\right)}
\def\[{\left[}
\def\]{\right]}

\def\e{\begin{equation}}
\def\q{\end{equation}}
\def\m{\begin{eqnarray}}
\def\n{\end{eqnarray}}
\newcommand{\Mov}[1]{{\color{black}{#1}}}
\newcommand{\Ke}[1]{{\color{black}{#1}}}

\begin{document}

\title{Implicit Likelihood Inference of the Neutrino Mass Hierarchy from Cosmological Data} 

\author{Ke Wang$^{1}$}
\thanks{{wangke@lnnu.edu.cn}}
\author{Jia-Yi Feng$^{1}$}
\affiliation{$^1$Department of Physics, Liaoning Normal University, Dalian 116029, China}

\date{\today}

\begin{abstract}
In this paper, we turn to the Learning the Universe Implicit Likelihood Inference (LtU-ILI) pipeline to perform a multi-round ILI of the neutrino mass hierarchy from cosmological data, including $TT$, $TE$, $EE$ power spectra of Planck 2018 and distance ratios of DESI DR2. More precisely, we first embed the CMB power spectra simulator $\mathtt{CLASS}$ into the LtU-ILI pipeline. And then, opting for Sequential Neural Likelihood Estimation (SNLE), we sequentially train neural networks using \Mov{$6$} rounds of \Mov{$10000$} simulations to target a ``black box'' likelihood of our forward model with one additional neutrino mass hierarchy parameter $\tilde{\Delta}$ and six base cosmological parameters. We find $\tilde{\Delta}=0.12^{+0.21}_{-0.23}~(68\%{\rm CL})$.
\end{abstract}

\pacs{???}

\maketitle


\section{Introduction}
\label{sec:intro}
Neutrino oscillation experiments require that there are three distinct mass eigenstates with the differences of squared neutrino masses as~\cite{ParticleDataGroup:2024cfk}
\begin{align}
\Delta m^2_{21}&=m^2_{2}-m^2_{1}=7.5\times10^{-5}{\rm eV^2},\\
|\Delta m^2_{31}|&=|m^2_{3}-m^2_{1}|=2.45\times10^{-3}{\rm eV^2}.
\end{align}
Since all neutrino mass eigenstates are supposed to be sufficiently light \Mov{$<0.45{\rm eV}$~\cite{KATRIN:2024cdt}}, these massive cosmic neutrinos would be relativistic and behave as the radiation in the early universe, while they would change to be non-relativistic and contribute to the matter budget in the late universe. This transition would leave footprints in the cosmological expansion history~\cite{Lesgourgues:2006nd}. Furthermore, also in the late universe, the large thermal velocities of non-relativistic neutrinos would suppress the growth of structure below the neutrino free-streaming scale~\cite{Lesgourgues:2006nd}.
Therefore, the combination of the cosmic microwave background (CMB, Planck 2018~\cite{Planck:2018vyg}), the baryon acoustic oscillation (BAO, DESI DR2~\cite{DESI:2025zgx}) and the matter power spectrum~\cite{Chabanier:2019eai} can impose a tight constraint on the neutrino masses, hence the neutrino mass hierarchy.
For example, the combination of Planck, ACT~\cite{AtacamaCosmologyTelescope:2025blo} and DESI DR2 provides a strong constraint, $\sum m_{\nu}<0.077{\rm eV}~(95\%{\rm CL})$~\cite{DESI:2025gwf}. This constraint becomes excessively tighter by including recent SPT-3G data~\cite{SPT-3G:2025bzu}, $\sum m_{\nu}<0.048{\rm eV}~(95\%{\rm CL})$. Such strong bounds are consistent with the preference for the normal hierarchy $m_1<m_3$ from~\cite{Huang:2015wrx,RoyChoudhury:2019hls,Jiang:2024viw,Ivanov:2026dvl}.

To implement the inference of the neutrino masses and other cosmological parameters from observations by the traditional explicit likelihood inference, even for CMB data from Planck 2018~\cite{Planck:2018vyg} only, one should turn to the low-$\ell$ temperature-only $\mathtt{Commander}$ likelihood to deal with the temperature power spectrum in the multipole range $2 \leq \ell \leq 29$, the low-$\ell$ EE likelihood from $\mathtt{SimAll}$ to deal with $EE$-polarization power spectrum in the multipole range $2 \leq \ell \leq 29$, the $\mathtt{Plik}$ high-multipole likelihood to deal with $TT$, $TE$ and $EE$ angular power spectra at multipoles $\ell\geq 30$ and the CMB lensing likelihood to deal with the CMB lensing-potential power spectrum, respectively. 
The above 4 likelihoods only represent \Mov{certain} summary statistics of CMB anisotropies map. \Mov{In fact, there are some other summary statistics, such as the temperature and polarization angular bispectra. The more summary statistics are considered, the more nuisance parameters should be introduced. Furthermore, $6+n$ additional cosmological parameters must be constrained at the same time when an extension to the base-$\Lambda$CDM model is studied. However, the traditional explicit likelihood inference scales poorly with the dimensionality of parameter space.}

Implicit Likelihood Inference (ILI)~\cite{Cranmer:2019eaq}, also known as simulation-based inference (SBI) and likelihood-free inference (LFI), is a novel alternative that can learn the statistical relationship between parameters and data by machine learning (ML). 
\Mov{Compared to the traditional explicit likelihood inference, instead of sampling the full joint posterior first and marginalizing it then, ILI can directly learn the marginal posteriors.}
Due to the prohibitive computational cost of $N$-body simulation~\cite{Castorina:2015bma,Liu:2017now,Villaescusa-Navarro:2019bje}, the training data in question are usually obtained before ILI and compressed to summary statistics, such as power spectrum~\cite{Hahn:2022wgo}, bispectrum~\cite{SimBIG:2023nol}, marked power spectrum~\cite{Massara:2024cvu}, skew spectrum~\cite{Hou:2024blc}, wavelet statistics~\cite{SimBIG:2023gke} and field-level statistics~\cite{SimBIG:2023ywd}.
However, $\mathtt{CLASS}$~\cite{Blas:2011rf} and $\mathtt{CAMB}$~\cite{Lewis:1999bs} as CMB simulators are efficient. Therefore, given an initial wider cosmological parameter priors, these CMB simulators can forward-simulate sets of parameter values to the training data of $TT$, $TE$, $EE$ and CMB lensing-potential power spectra. After the first round of training, the obtained posteriors serve as the new priors for the second round of training, and so on. This sequential learning has been applied to ILI of the cosmological parameters from CMB~\cite{Cole:2021gwr}.

ILI of neutrino masses from BOSS voids obtains a modest constraint on $\sum m_{\nu}$~\cite{Thiele:2023oqf}.
\Mov{In this paper, we expect a strong limit on $\sum m_{\nu}$ from the combination of DESI DR2~\cite{DESI:2025zgx} and Planck 2018~\cite{Planck:2018vyg} and perform ILI of the neutrino mass hierarchy by sequential learning.}
To study the preference for what neutrino mass hierarchy directly, we extend the base-$\Lambda$CDM model with the neutrino mass hierarchy parameter~\cite{Jimenez:2010ev,Xu:2016ddc}
\begin{equation}
\Delta=\frac{m_3-m_1}{m_1+m_3},
\end{equation}
where $\Delta>0$ for the normal hierarchy and $\Delta<0$ for the inverted hierarchy respectively.

This paper is organized as follows. In Section~\ref{sec:simulator}, we introduce the CMB power spectra simulator including the neutrino mass hierarchy parameter. In Section~\ref{sec:ili}, we perform a multi-round ILI of the neutrino mass hierarchy. Finally, a brief summary and discussion are provided in Section~\ref{sec:sum}.

\section{CMB power spectra simulator}
\label{sec:simulator}
\begin{figure*}[]
\begin{center}
\includegraphics[width= 8cm]{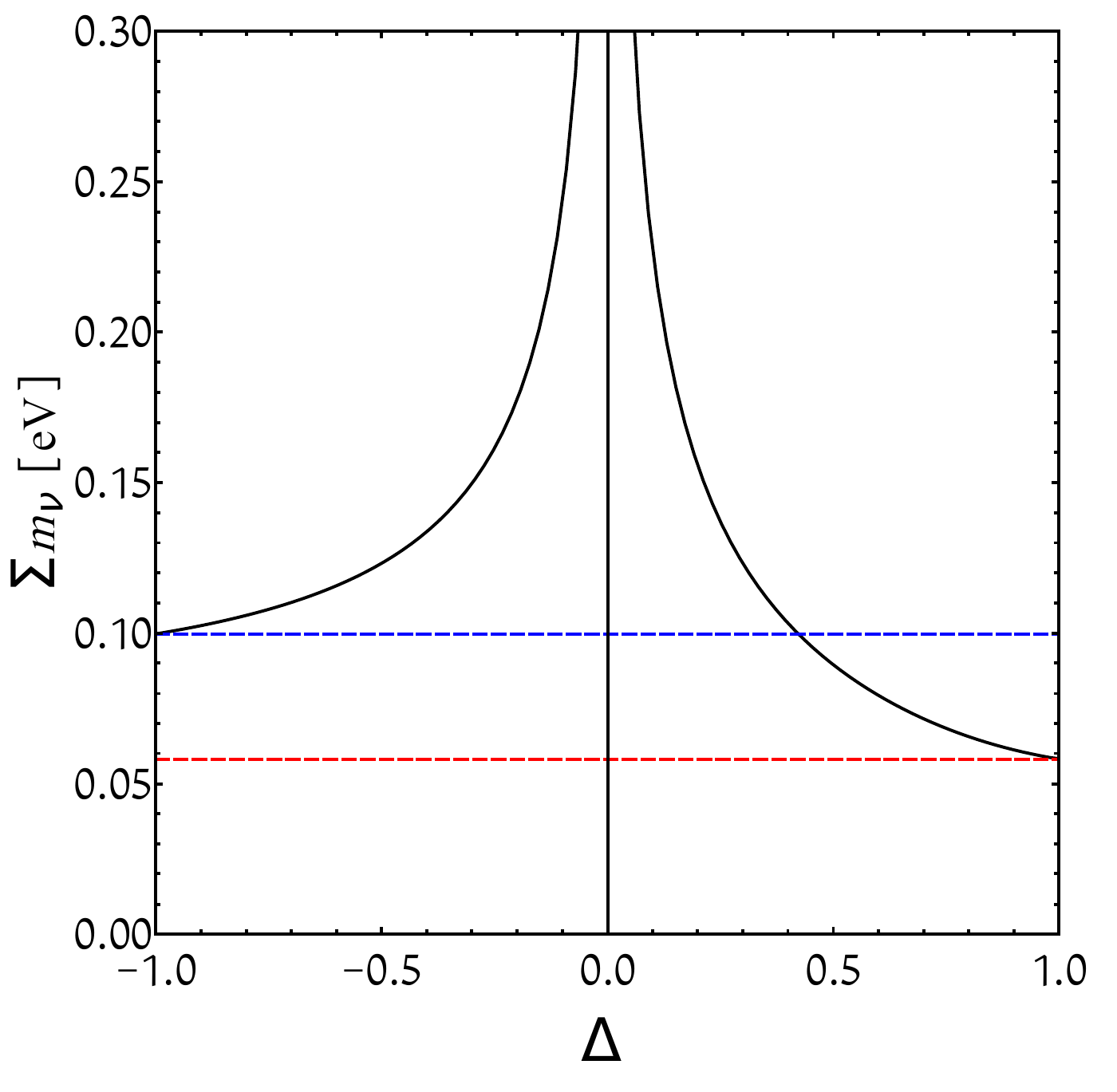}
\includegraphics[width= 8cm]{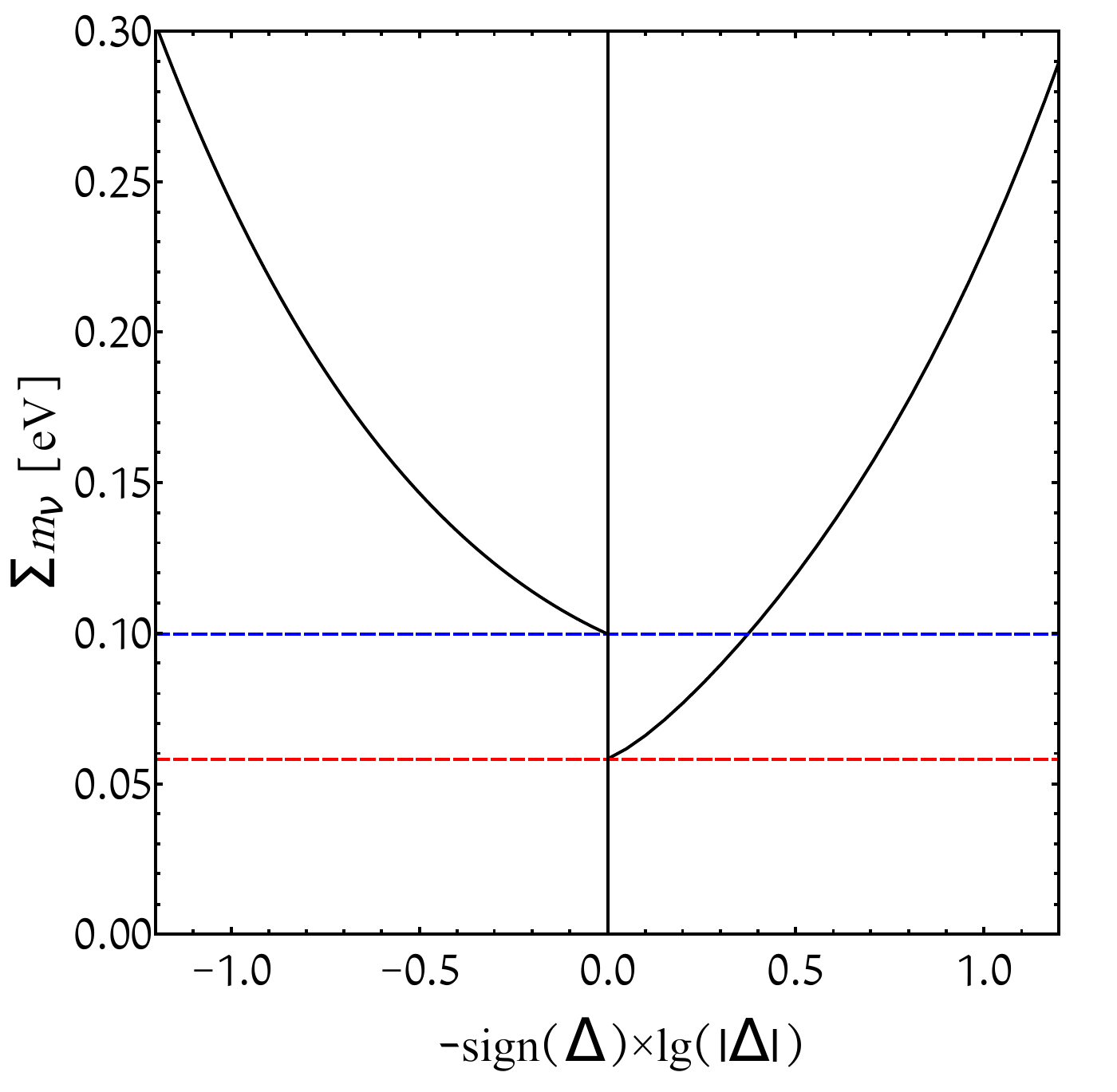}
\end{center}
\caption{The total neutrino mass $\sum m_{\nu}$ as a function of $\Delta$ (left) or $\tilde{\Delta}=-{\rm sign}(\Delta)\times\lg(|\Delta|)$ (right), according to which an upper limit on $\sum m_{\nu}$ can be derived from a $\cup$-shaped posterior of $\Delta$ or a $\cap$-shaped posterior of $\tilde{\Delta}$.}
\label{fig:delta}
\end{figure*}

The neutrino mass hierarchy parameter $\Delta$ is related to the total neutrino mass as~\cite{Jimenez:2010ev,Xu:2016ddc}
\begin{equation}
\sum m_{\nu}=\sqrt{\frac{|\Delta m^2_{31}|}{|\Delta|}}+\sqrt{\frac{(1-\Delta)^2}{4|\Delta|}|\Delta m^2_{31}|+\Delta m^2_{21}}.
\end{equation}
Therefore, the usual upper limits on $\sum m_{\nu}$ mean that $\Delta$ should have a $\cup$-shaped distribution, as hinted at by the left subplot of Fig.~\ref{fig:delta}. This total non-Gaussian $\cup$-shaped posterior maybe can not be estimated from the data correctly.
Here, we introduce a new parameter $\tilde{\Delta}=-{\rm sign}(\Delta)\times\lg(|\Delta|)$ which is related to $\sum m_{\nu}$ as the right subplot of Fig.~\ref{fig:delta}. Consequently, an upper limit on $\sum m_{\nu}$ would be equivalent to a $\cap$-shaped posterior of $\tilde{\Delta}$. Taking $\tilde{\Delta}$ into account, our cosmological model has $7$ free parameters. Since we will turn to the sequential learning, their initial uniform priors can be wider
\begin{equation}
\label{eq:prior} 
\bm{\theta}=\left\{
\begin{aligned}
&\omega_{\rm b}\\
&\omega_{\rm c}\\
&100\theta_s\\
&\ln (10^{10}A_{\rm s})\\
&n_{\rm s}\\
&\tau\\
&\tilde{\Delta}
\end{aligned}
\right\}= 
\left\{
\begin{aligned}
&\mathcal{U}(0.020,0.024)\\
&\mathcal{U}(0.11,0.13)\\
&\mathcal{U}(1.036,1.044)\\
&\mathcal{U}(2.98,3.20)\\
&\mathcal{U}(0.94,1.02)\\
&\mathcal{U}(0.04,1.00)\\
&\mathcal{U}(-3,3)
\end{aligned}
\right\}.
\end{equation}
\Mov{For $6$ base cosmological parameters, above priors are wider than $\pm5\sigma$ ranges of constraints from the combination of Planck 2018 and BAO~[3] by the traditional explicit likelihood inference; for $\tilde{\Delta}$, its prior allows for $\sum m_{\nu}>2{\rm eV}$}.

\begin{figure*}[]
\begin{center}
\includegraphics[width= 11cm]{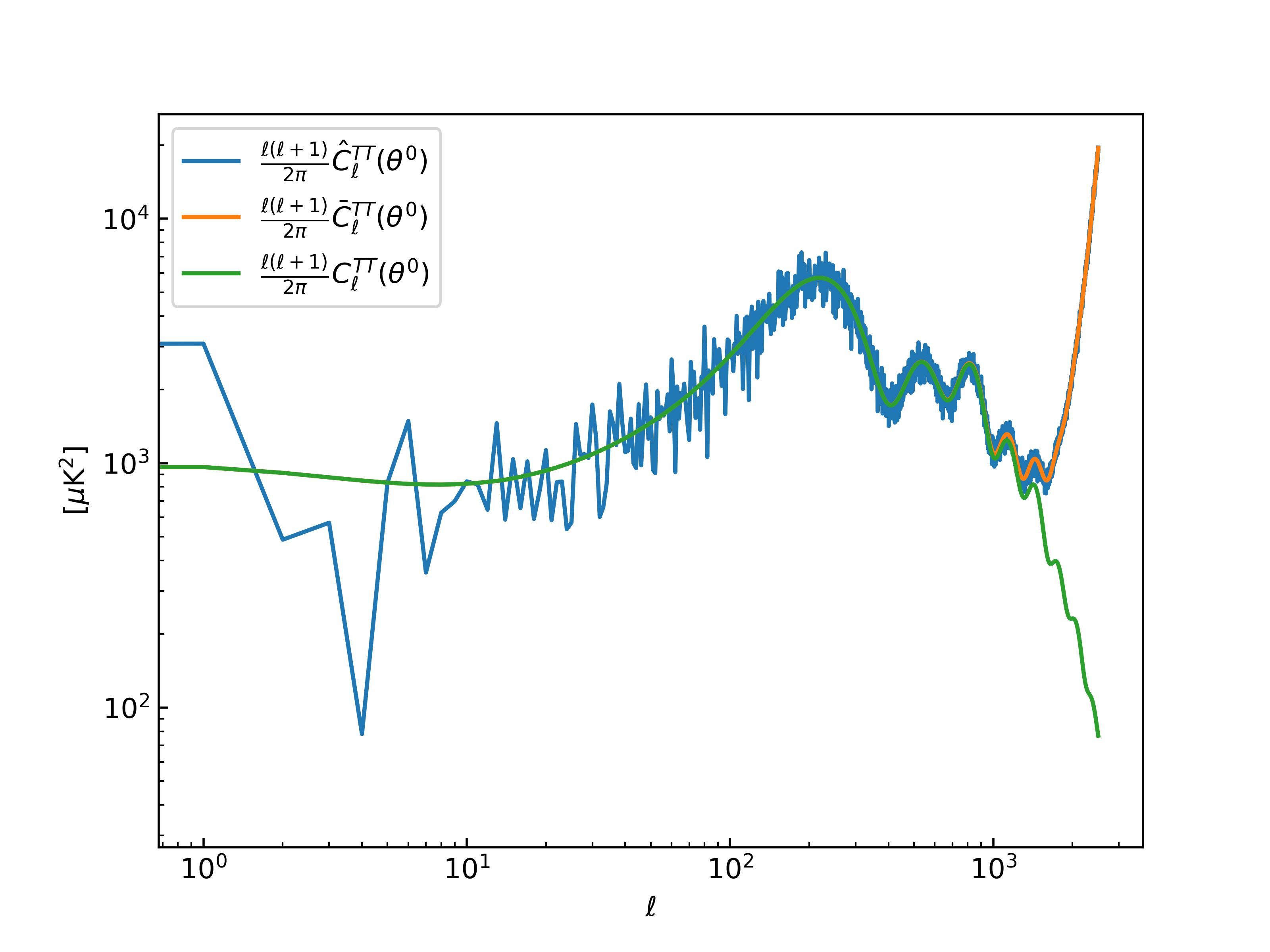}\\
\includegraphics[width= 11cm]{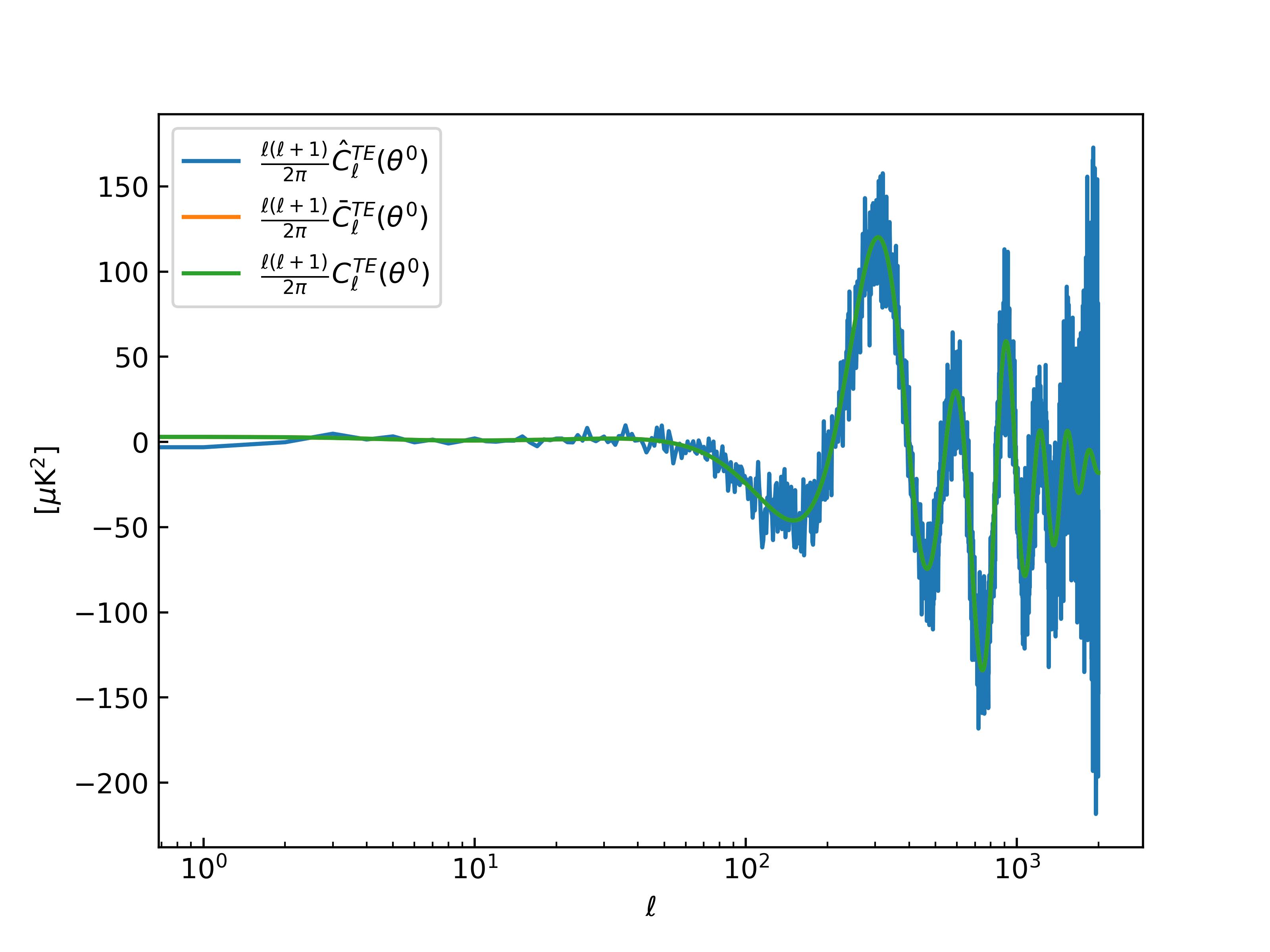}\\
\includegraphics[width= 11cm]{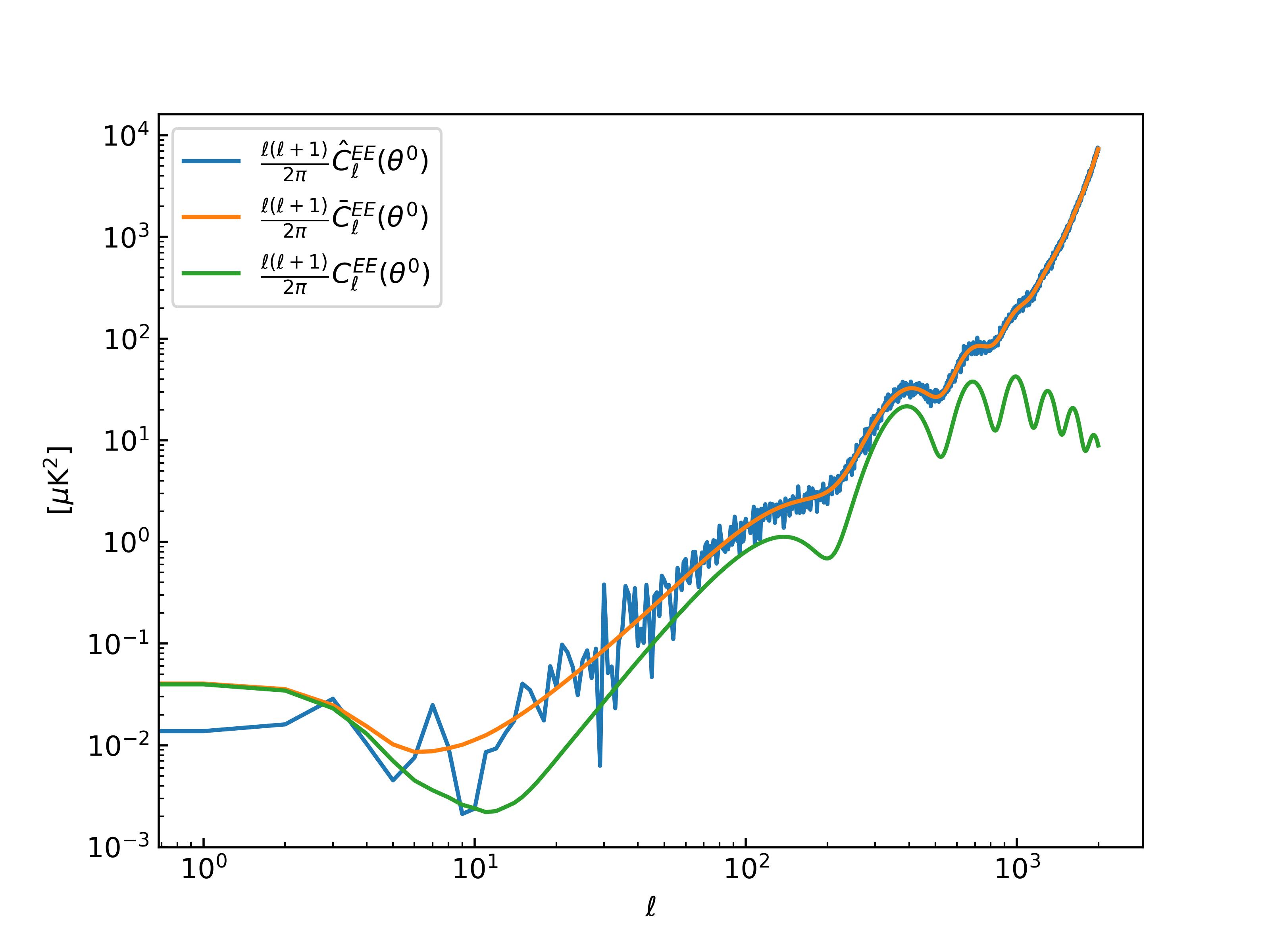}
\end{center}
\caption{Comparison among $C_{\ell}^{XX'}(\bm{\theta}^0)$, $\bar{C}_{\ell}^{XX'}(\bm{\theta}^0)$ and $\hat{C}_{\ell}^{XX'}(\bm{\theta}^0)$ respectively, where $\bm{\theta}^0$ does not include the neutrino mass hierarchy parameter $\tilde{\Delta}$.}
\label{fig:Cls}
\end{figure*}

As the procedure of~\cite{Cole:2021gwr}, the CMB simulator is constructed by three steps. 1) Given the prior $\mathcal{P}(\bm{\theta})$, we can draw sets of candidate parameter values from it and forward-simulate these values to the corresponding CMB power spectra $C_{\ell}^{XX'}(\bm{\theta})$ ($X=T,E$) by the Einstein-Boltzmann solver $\mathtt{CLASS}$~\cite{Blas:2011rf}. 2) For the Planck experiment, its noise spectrum can be mimicked by the contributions of a Gaussian white noise and a Gaussian beam~\cite{CORE:2016npo}
\begin{equation}
N_{\ell}^{XX'}=\delta_{XX'}\theta_{\rm FWHM}^2\sigma_X^2\exp\left(\ell(\ell+1)+\frac{\theta_{\rm FWHM}^2}{8\ln2}\right),
\end{equation}
where $\theta_{\rm FWHM}$ is the full width at half maximum of a Gaussian beam and $\sigma_X$ is the detector sensitivity. Thanks to $\mathtt{fake\_planck\_realistic}$ within $\mathtt{MontePython}$~\cite{Audren:2012wb}, a ready-made $N_{\ell}^{XX'}$ for Planck's temperature and polarization measurements, we have $\bar{C}_{\ell}^{XX'}(\bm{\theta})=C_{\ell}^{XX'}(\bm{\theta})+N_{\ell}^{XX'}$. 3) Given the covariance matrix 
\begin{equation}
\label{eq:covmatrix}
\left(
\begin{aligned}
&\bar{C}_{\ell}^{TT}(\bm{\theta})~~\bar{C}_{\ell}^{TE}(\bm{\theta})\\
&\bar{C}_{\ell}^{TE}(\bm{\theta})~~\bar{C}_{\ell}^{EE}(\bm{\theta})
\end{aligned}
\right),
\end{equation}
the Gaussian distributions of the multipole coefficients $a_{\ell m}^{X}$ of CMB maps can be sampled accordingly.
For a specific couple of $\ell'$ and $m'$, if the corresponding covariance matrix block is decomposed into the product of a lower triangular matrix $L$ and its conjugate transpose $L^T$ using the Cholesky decomposition, $a_{\ell' m'}^{X}$ can be sampled as 
\begin{equation}
\left(
\begin{aligned}
&a_{\ell' m'}^{T}\\
&a_{\ell' m'}^{E}
\end{aligned}
\right)=L(\bm{\theta}) 
\left(
\begin{aligned}
&\mathcal{N}(0,1)\\
&\mathcal{N}(0,1)
\end{aligned}
\right).
\end{equation}
According to the estimator of power spectra
\begin{equation}
\hat{C}_{\ell}^{XX'}(\bm{\theta})=\frac{1}{2\ell+1}\sum_{m=-\ell}^{\ell}a_{\ell m}^{X*}a_{\ell m}^{X'},
\end{equation} 
there obviously is a fundamental uncertainty (due to cosmic variance) in the knowledge about the power spectra.
More precisely, $\hat{C}_{\ell}^{XX'}(\bm{\theta})$ has a Wishart distribution at low $\ell$ and tends towards a multivariate Gaussian distribution at large $\ell$~\cite{Percival:2006ss}. 
For simplicity, at $\ell>52$, we set $\hat{C}_{\ell}^{XX'}(\bm{\theta})=\bar{C}_{\ell}^{XX'}(\bm{\theta})+n_{\ell}^{XX'}(\bm{\theta}^0)$, where $n_{\ell}^{XX'}(\bm{\theta}^0)$ depends on a fiducial cosmology\footnote{Here we directly download the theoretical power spectra of the best-fit model $C_{\ell}^{XX'}$ from Planck 2018~\cite{PLA}, which can also be reproduced by a specific set of parameters $\bm{\theta}^0=\{\omega_{\rm b},\omega_{\rm c},100\theta_s,\ln (10^{10}A_{\rm s}),n_{\rm s},\tau\}=\{0.02237,0.1200,1.04092,3.044,0.9649,0.0544\}$.} and can be simply sampled from the corresponding multivariate normal distribution with the covariance matrix
\begin{align}
\nonumber
&\frac{2}{2\ell+1}\times\\
&\left(
\begin{aligned}
&\left(\bar{C}_{\ell}^{TT}(\bm{\theta}^0)\right)^2~~~~~~~~~~\bar{C}_{\ell}^{TT}(\bm{\theta}^0)\bar{C}_{\ell}^{TE}(\bm{\theta}^0)~~~~~~~~~\left(\bar{C}_{\ell}^{TE}(\bm{\theta}^0)\right)^2\\
&~~~~~~~~~~~~~~~~~~~~~~~~~~~~~~~~\left.\frac{1}{2}\right(\\
&\bar{C}_{\ell}^{TT}(\bm{\theta}^0)\bar{C}_{\ell}^{TE}(\bm{\theta}^0)~~\bar{C}_{\ell}^{TT}(\bm{\theta}^0)\bar{C}_{\ell}^{EE}(\bm{\theta}^0)~~\bar{C}_{\ell}^{TE}(\bm{\theta}^0)\bar{C}_{\ell}^{EE}(\bm{\theta}^0)\\
&~~~~~~~~~~~~~~~~~~~~~~~~~+\left.\left(\bar{C}_{\ell}^{TE}(\bm{\theta}^0)\right)^2\right)\\
&\left(\bar{C}_{\ell}^{TE}(\bm{\theta}^0)\right)^2~~~~~~~~~~\bar{C}_{\ell}^{TE}(\bm{\theta}^0)\bar{C}_{\ell}^{EE}(\bm{\theta}^0)~~~~~~~~~\left(\bar{C}_{\ell}^{EE}(\bm{\theta}^0)\right)^2
\end{aligned}
\right).
\end{align}
In Fig.~\ref{fig:Cls}, we compare $C_{\ell}^{XX'}(\bm{\theta}^0)$, $\bar{C}_{\ell}^{XX'}(\bm{\theta}^0)$ and $\hat{C}_{\ell}^{XX'}(\bm{\theta}^0)$ respectively.

\Mov{As for the CMB lensing simulator, given the $9\times9$ covariance matrix $\mathtt{smi...\_ndclpp\_p\_teb\_consext8\_cov}$ of Planck 2018 minimum-variance (MV) lensing band powers at multipole range $8\sim400$~\cite{Planck:2018lbu,PLA}, the noise realizations of the lensing power spectrum at $9$ weighted multipole-bin centres are sampled from the corresponding multivariate normal distribution.}
\Mov{Similarly,} for the BAO simulator, given the covariance matrix of distance ratios of DESI DR2~\cite{DESI:2025zgx}, the noise realizations are sampled from the corresponding multivariate normal distribution.

\section{Multi-round ILI of the neutrino mass hierarchy}
\label{sec:ili}
In this paper, we turn to the Learning the Universe Implicit Likelihood Inference ($\mathtt{LtU}$-$\mathtt{ILI}$) pipeline~\cite{Ho:2024whi}, where the computational backend is set as $\mathtt{sbi}$~\cite{sbi}, a $\mathtt{PyTorch}$~\cite{PyTorch} package, and the sequential analog of Neural Likelihood Estimation (SNLE)~\cite{Alsing:2018eau,Papamakarios:2019ccu} is chosen as the inference engine.
The loss function for NLE is 
\begin{equation}
\mathcal{L}=-\mathbb{E}_{\mathcal{D_{\rm train}}}[\ln q_{\bm w}({\bm x}|{\bm \theta})],
\end{equation}
where ${\bm w}$ is the network weights, the expectation $\mathbb{E}_{\mathcal{D_{\rm train}}}$ is taken over all data-parameter pairs $\{({\bm x}_i,{\bm \theta}_i)\}$ in the training data set $\mathcal{D_{\rm train}}$ and $q_{\bm w}({\bm x}|{\bm \theta})$ is converging to the likelihood through the minimization of $\mathcal{L}$ over ${\bm w}$. 
Then, we sequentially perform \Mov{$6$} rounds of training. In each round, we train an NLE model with \Mov{$10000$} simulated data-parameter pairs $\{({\bm x}_i,{\bm \theta}_i)\}$ by an ensemble of \Mov{$6$} Neural Density Estimations~\cite{Papamakarios:2019ccu} (NDEs), each using a Masked Autoregressive Flow~\cite{MAF} (MAF) architecture with $50$ hidden features and $5$ transformations.

After the last round of training, the learned likelihood is related to an amortized posterior as $\mathcal{P}({\bm \theta}|{\bm x})\propto q_{\bm w}({\bm x}|{\bm \theta})\mathcal{P}({\bm \theta})$. 
Due to the local amortization of $\mathcal{P}({\bm \theta}|{\bm x})$, the marginal posterior of $\bm\theta$ can be sampled from $\mathcal{P}({\bm \theta}|{\bm x})$ without retraining, for \Mov{a specific} $\bm x$.
Moreover, local amortization allows for special validation tests that are not accessible to traditional explicit likelihood inference, such as the rank statistics, the marginal percentile coverage test and the direct comparison between true and predicted parameter values.
Similarly to the training data set $\mathcal{D_{\rm train}}$, we prepare \Mov{$1000$} data-parameter pairs $\{({\bm x}_i,{\bm \theta}_i)\}$ as the testing data set $\mathcal{D_{\rm test}}$. $\mathcal{D_{\rm test}}$ was not used in the SNLE training, but their marginal posteriors are sampled according to the local amortization of $\mathcal{P}({\bm \theta}|{\bm x})$ for the following validation.

\begin{figure*}[]
\begin{center}
\includegraphics[width= 18cm]{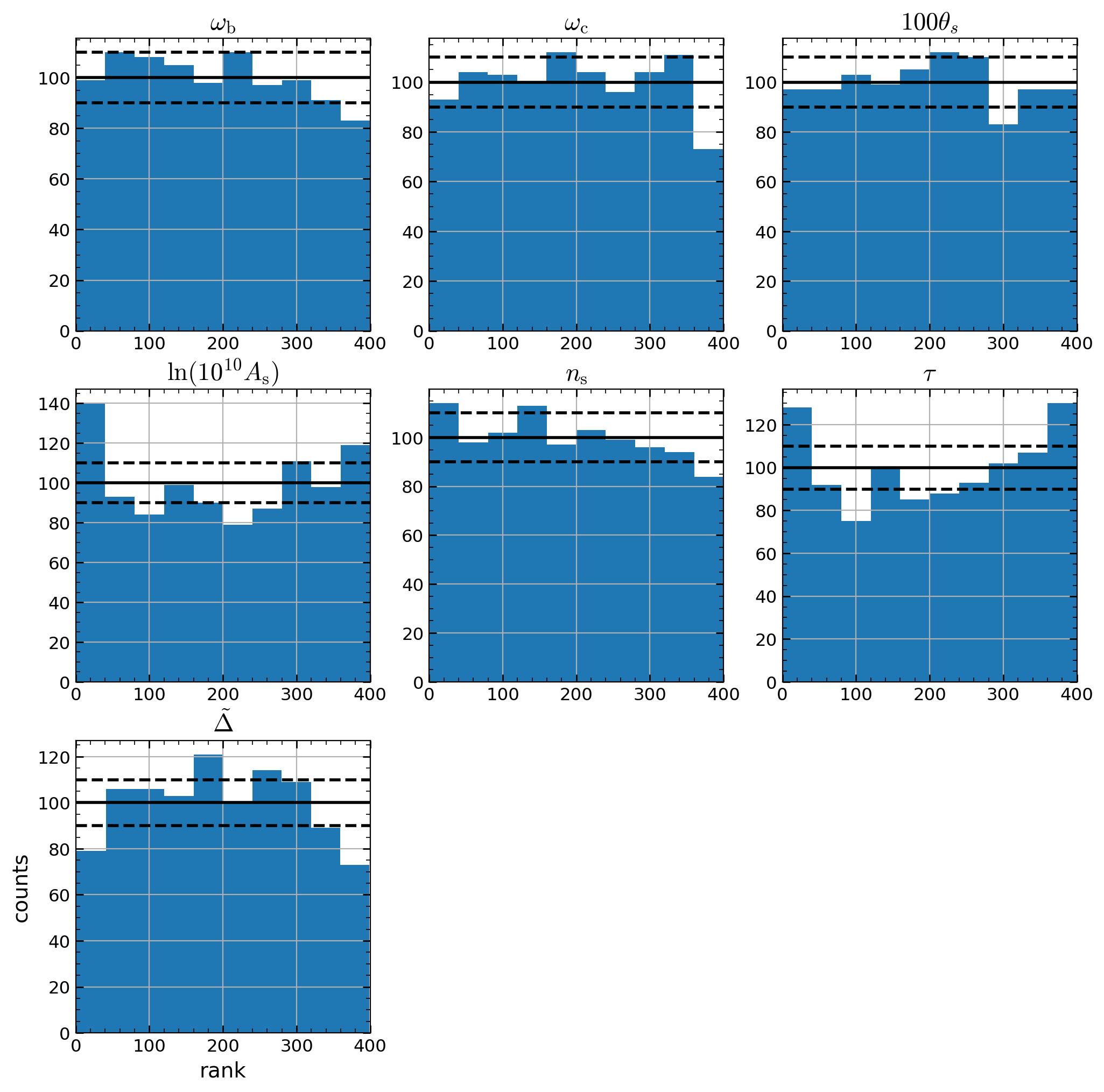}
\end{center}
\caption{Rank statistic of each parameter, \Mov{where $\sum {\rm counts} = 1000$ is the number of data-parameter pairs in the testing data set $\mathcal{D_{\rm test}}$ and the rank range of $0$ to $400$ results from the comparison between ${\theta}_i$ and its $N=400$ posterior samples drawn from $\mathcal{P}({\theta}|{\bm x}_i)$ for a specific data-parameter pair $({\bm x}_i,{\bm \theta}_i)$ from $\mathcal{D_{\rm test}}$.
The distribution and deviation from uniformity of rank statistic can serve as the diagnostics for the incorrectness of posterior~\cite{Hahn:2022wgo,Talts:2018zdk,Hahn:2022nda}: 
if we estimate $\mathcal{P}({\bm \theta}|{\bm x})$ exactly, the rank statistics would be distributed uniformly, as the black solid lines;
while an $\cup$-shaped rank statistic means that our estimated $\mathcal{P}({\bm \theta}|{\bm x})$ is narrower than the true posterior, a $\cap$-shape rank statistic means that our estimated $\mathcal{P}({\bm \theta}|{\bm x})$ is broader than the true posterior;
an asymmetric rank histogram implies that our estimated $\mathcal{P}({\bm \theta}|{\bm x})$ is biased in the opposite direction relative to the true posterior;
the spikes at the boundaries of histogram means there are obvious correlations between parameters.}}
\label{fig:rank}
\end{figure*}

\begin{figure*}[]
\begin{center}
\includegraphics[width= 18cm]{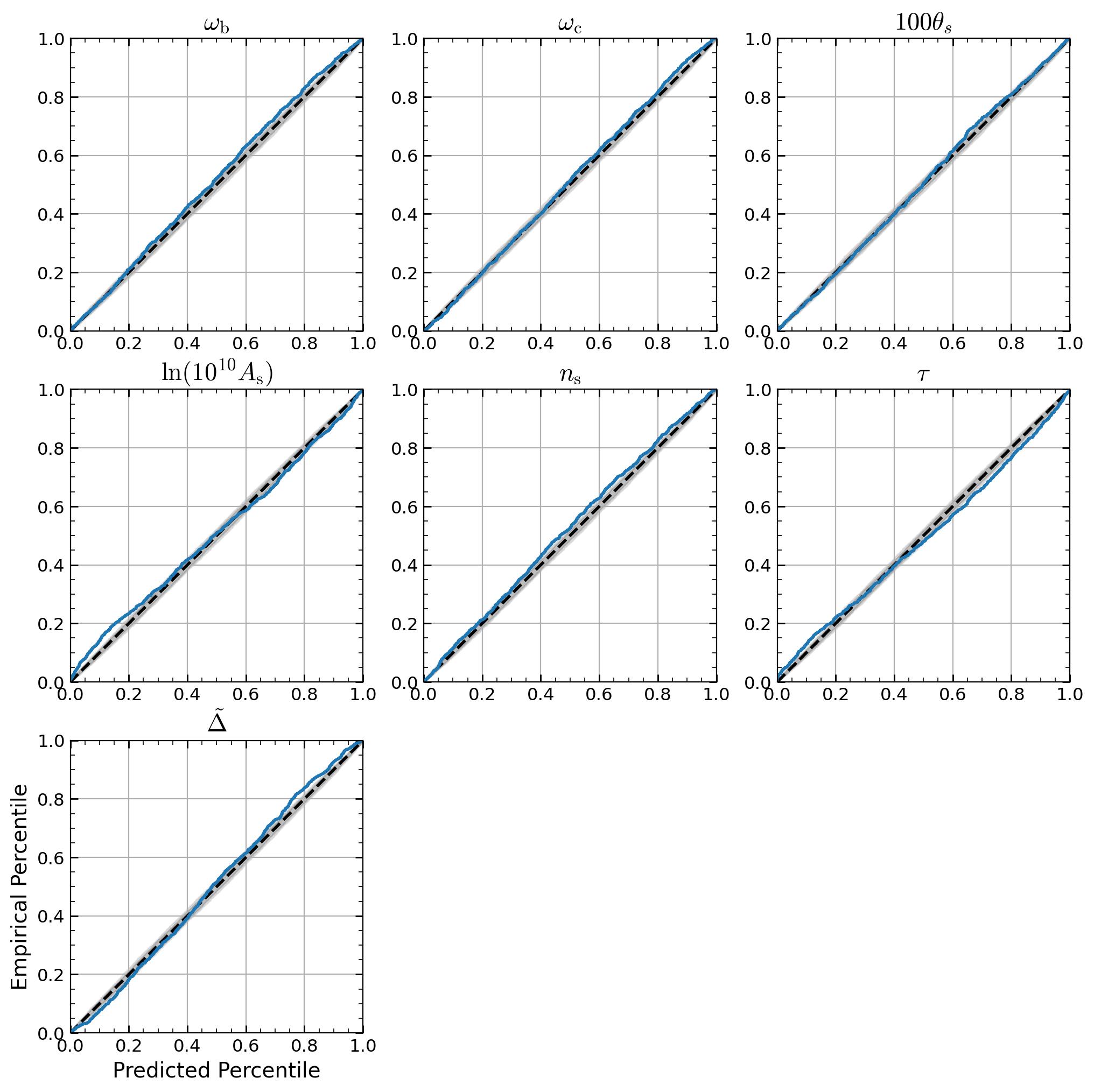}
\end{center}
\caption{Marginal percentile coverage test for cosmological parameters, where the empirical percentile from uniform distribution $\mathcal{U}(0,1)$ and the predicted percentile from the distribution of all CDF values calculated at $({\bm x}_i,{\bm \theta}_i)\in\mathcal{D_{\rm test}}$ as Eq.~(\ref{eq:CDF}).}
\label{fig:pp}
\end{figure*}

\begin{figure*}[]
\begin{center}
\includegraphics[width= 18cm]{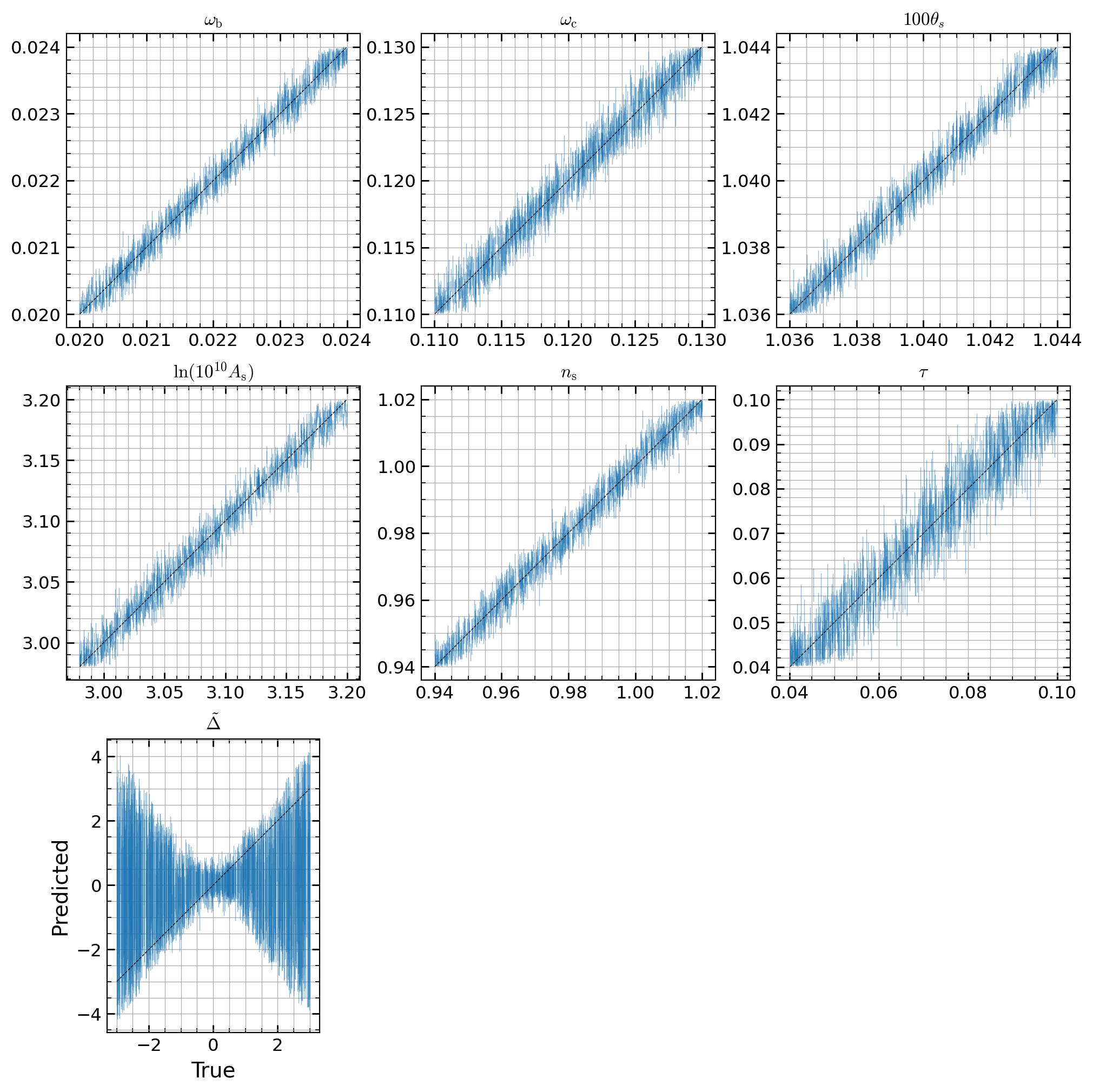}
\end{center}
\caption{Comparison between true and predicted value of cosmological parameters.}
\label{fig:TvsP}
\end{figure*}

Given a data-parameter pair $({\bm x}_i,{\bm \theta}_i)$ from $\mathcal{D_{\rm test}}$, we can count the rank of each given parameter $\theta_i$ in its $N$ posterior samples
\begin{equation}
\{\theta'_1,\theta'_2,...,\theta'_{N=400}\}\sim\mathcal{P}({\theta}|{\bm x}_i),~\forall ({\bm x}_i,{\bm \theta}_i)\in\mathcal{D_{\rm test}}.
\end{equation}
\Mov{For example, $\theta_i$ has rank $3$ when $\theta'_2<\theta_i<\theta'_3$.} 
If the true posterior is estimated exactly, the rank statistic of $\theta_i$ in $\mathcal{D_{\rm test}}$ should be distributed uniformly~\cite{Hahn:2022wgo,Talts:2018zdk,Hahn:2022nda}. This is almost the case for our cosmological parameters, as shown in Fig.~\ref{fig:rank}.
\Mov{However, there are two obvious left spikes for $\ln(10^{10}A_{\rm s})$ and $\tau$, which results from the correlation between them. Furthermore, a subtle $\cap$-shape rank statistic for $\tilde{\Delta}$ means that our estimated $\mathcal{P}(\tilde{\Delta}|{\bm x})$ is slightly broader than the true one.}
Replacing the rank with the cumulative density function (CDF) value, the distribution of the latter one would also be uniformly distributed
\begin{equation}
\label{eq:CDF}
\int_{-\infty}^{\theta_i}d\theta\mathcal{P}({\theta}|{\bm x}_i)\sim\mathcal{U}(0,1),~\forall ({\bm x}_i,{\bm \theta}_i)\in\mathcal{D_{\rm test}}.
\end{equation}
In the percentile-percentile (P-P) plot, Fig.~\ref{fig:pp}, we compare the CDF of the distribution of all CDF values calculated at $({\bm x}_i,{\bm \theta}_i)\in\mathcal{D_{\rm test}}$ with the CDF of $\mathcal{U}(0,1)$. The good agreement between each couple of distributions means that our learned posteriors are almost globally consistent with the truth.
Finally, in Fig.~\ref{fig:TvsP}, we directly compare the true value of $\theta_i$ in $\mathcal{D_{\rm test}}$ and the corresponding predicted values sampled from $\mathcal{P}({\theta}|{\bm x}_i)$. For a large $|\tilde{\Delta}|$, or a large $\sum m_{\nu}$ according to Fig.~\ref{fig:delta}, cosmological data can not distinguish the normal hierarchy and the inverted hierarchy. So given a large true $|\tilde{\Delta}|$, its predicted one can be either $>0$ or $<0$, as shown in the last subplot in Fig.~\ref{fig:TvsP}. As $|\tilde{\Delta}|$ is decreasing, the hierarchy preference occurs.

\begin{figure*}[]
\begin{center}
\includegraphics[width= 18cm]{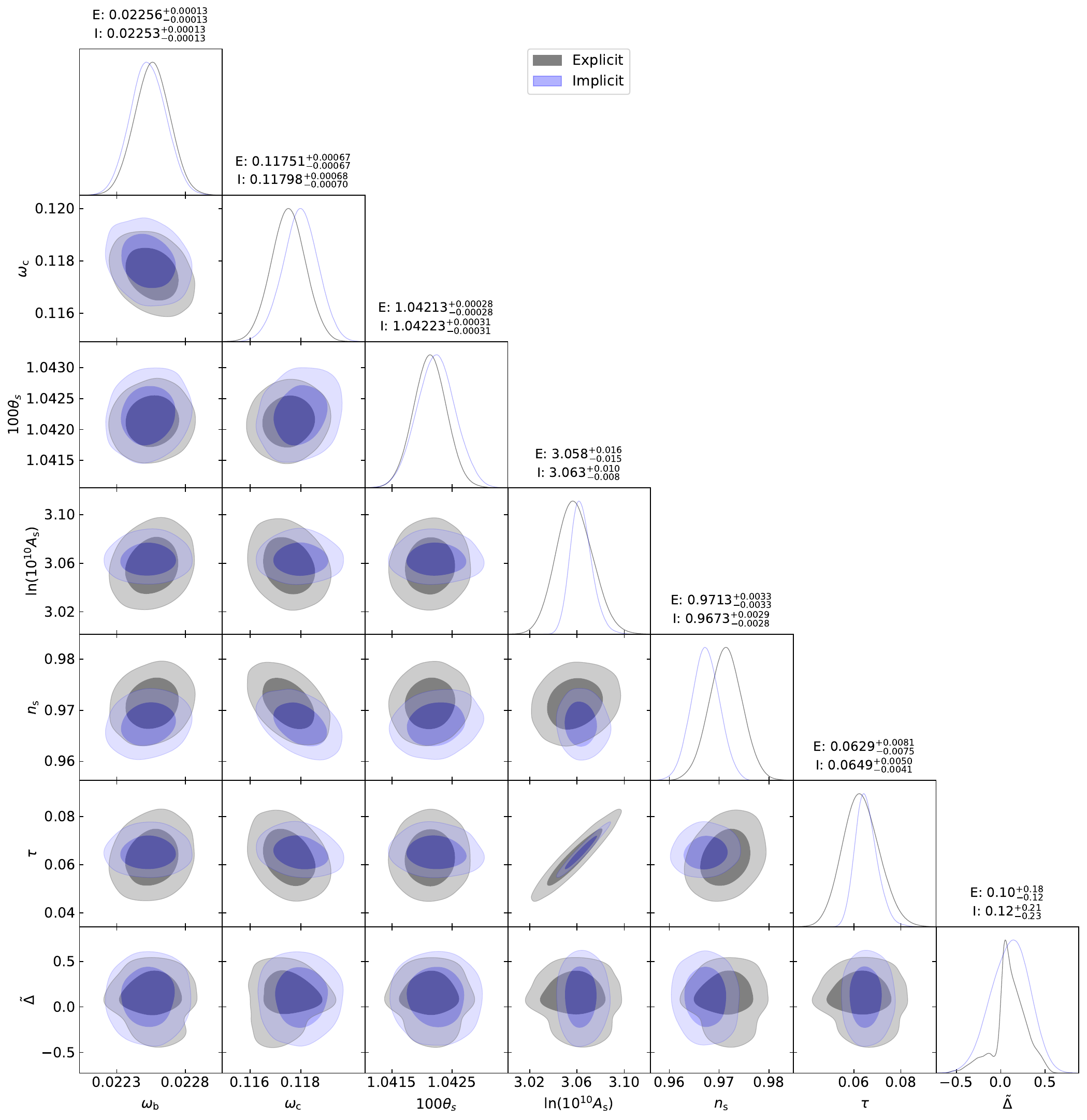}
\end{center}
\caption{Posterior distribution of $\bm{\theta}$ inferred from the combination
of DESI DR2~\cite{DESI:2025zgx} and Planck 2018~\cite{Planck:2018vyg} \Mov{by ILI (blue) and the traditional explicit likelihood inference (gray)}, where $68\%$ interval of each parameter is displayed as the titles of diagonal subplots and contours contain $68\%$ and $95\%$ of the probability.}
\label{fig:posterior}
\end{figure*}

After validation by $\mathcal{D_{\rm test}}$, we now consider the observed cosmological data $\bm{x}_o$, including the observed CMB \Mov{$TT$, $EE$ and $TE$} spectra and their uncertainties $\mathtt{COM\_PowerSpect\_CMB\_XX'}$ from Planck 2018~\cite{PLA}, \Mov{the CMB MV lensing band powers and their uncertainties $\mathtt{smi...\_ndclpp\_p\_teb\_consext8\_bandpowers}$ from Planck 2018~\cite{PLA}} and the distance ratios and their covariance matrix of DESI DR2~\cite{DESI:2025zgx}.
Given the amortized posterior $\mathcal{P}({\bm \theta}|{\bm x})$, the posterior \Mov{of $\bm \theta$} for $\bm{x}_o$ can then be obtained through the usual Monte Carlo Markov Chain sampling\Mov{. Here} we perform \Mov{it} with $\mathtt{emcee}$~\cite{Foreman-Mackey:2012any}. 
\Mov{
We also provide the constraints on these parameters from the same data combination by the traditional explicit likelihood inference, with $\mathtt{CLASS}$~\cite{Blas:2011rf} and $\mathtt{MontePython}$~\cite{Brinckmann:2018cvx}.
We compare the results from two methods in Fig.~\ref{fig:posterior}. We find that the constraints on all parameters obtained by ILI are well consistent with ones given by the traditional explicit likelihood inference, except for around 1-$\sigma$ difference between the constraints on $n_s$. 
More precisely, the constraint on the neutrino mass hierarchy parameter obtained by ILI is $\tilde{\Delta}=0.12^{+0.21}_{-0.23}~(68\%{\rm CL})$, whose probability ratio of $\tilde{\Delta}>0$ to $\tilde{\Delta}<0$ is $2.2:1$ after the integration of its probability density distribution; the constraint on $\tilde{\Delta}$ given by the traditional explicit likelihood inference is $\tilde{\Delta}=0.10^{+0.18}_{-0.12}~(68\%{\rm CL})$ which is a non-Gaussian distribution totally and its probability ratio of $\tilde{\Delta}>0$ to $\tilde{\Delta}<0$ is $4.5:1$.}
\Ke{As shown in the above validation tests including the rank statistic (Fig.~\ref{fig:rank}) and the marginal percentile coverage test (Fig.~\ref{fig:pp}), our estimated posteriors of some parameters don't exactly fit their true counterparts. This slight deviation leads to the less than 1-$\sigma$ difference between two methods' constraints shown in Fig.~\ref{fig:posterior}.
}

\section{Summary and discussion}
\label{sec:sum} 
In this paper, we first construct the CMB power spectra simulator $\hat{C}_{\ell}^{XX'}(\bm{\theta})$ by decorating the $\mathtt{CLASS}$ outputs $C_{\ell}^{XX'}(\bm{\theta})$ with the noise realizations of Planck experiment and the fundamental uncertainties due to cosmic variance.
Then, SNLE is chosen as the inference engine to target an amortized posterior $\mathcal{P}({\bm \theta}|{\bm x})$ indirectly.
After training neural networks with $\mathcal{D_{\rm train}}$,
we perform the validation test with the rank statistics, the P-P plot and the percentile coverage plot of $\mathcal{D_{\rm test}}$.
Given the credible learned $\mathcal{P}({\bm \theta}|{\bm x})$, $\mathcal{P}({\bm \theta}|{\bm x}_0)$ is sampled with $\mathtt{emcee}$~\cite{Foreman-Mackey:2012any}, where ${\bm x}_0$ stands for the the observed cosmological data including Planck 2018~\cite{PLA} and DESI DR2~\cite{DESI:2025zgx}. 
Finally, we find that $\tilde{\Delta}=0.12^{+0.21}_{-0.23}~(68\%{\rm CL})$.

As we know, $TT$, $EE$ and $TE$ angular power spectra \Mov{and the lensing-potential power spectrum~\cite{Planck:2018lbu}} are just \Mov{four} summary statistics of CMB anisotropies map. In fact, there exist other CMB summary statistics, such as the temperature and polarization angular bispectra and even the next higher-order non-Gaussianity (NG) correlation function, i.e., the CMB angular trispectra~\cite{Planck:2019kim}. In particular, embedding the simulators of the cross-correlations between CMB lensing and other large-scale structure observations~\cite{White:2021yvw,Verdiani:2025jgz} into ILI should impose stronger constraints on the neutrino masses and their hierarchy. We will do this in future work. For some summary statistics, their explicit likelihoods are ready-made or the construction of their simulators is difficult than the built of their explicit likelihoods. In these cases, the explicit likelihoods can be embedded into ILI through their effective simulators~\cite{FrancoAbellan:2025fkb}.

\vspace{5mm}
\noindent {\bf Acknowledgments}
We acknowledge the use of HPC Cluster of Tianhe II in National Supercomputing Center in Guangzhou.



\end{document}